\documentclass[final]{svjour2}
\usepackage{graphicx}
\usepackage{rotating}
\usepackage{amssymb}
\usepackage{mathptmx}
\usepackage[numbers]{natbib}
\makeatletter
\journalname{Journal of Low Temperature Physics}

\bibpunct{}{}{,}{s}{}{,}

\begin{document}

\newcommand{\hdblarrow}{H\makebox[0.9ex][l]{$\downdownarrows$}-}
\title{Sensitivity to Cosmic Rays of Cold Electron Bolometers for Space Applications}

\author{M. Salatino \and P. de Bernardis \and L.S. Kuzmin \and S. Mahashabde \and S. Masi}
\institute{M. Salatino \and P. de Bernardis \and S. Masi \at
              Sapienza University of Rome, Physics Department, p.le Aldo Moro, 5 00185 Rome, Italy\\
              Tel.: +39 06 4991 4342\\
              Fax: +39 06 4957697\\
              \email{maria.salatino@roma1.infn.it}            \\
           \and
           L.S. Kuzmin  \and S. Mahashabde\at
              Department of Microtechnology and Nanoscience - MC2 Chalmers University of Technology, SE-412 96 G\"{o}teborg, Sweden
}

\date{XX.XX.20XX}

\maketitle

\begin{abstract}
An important phenomenon limiting the sensitivity of bolometric detectors for future space missions is the interaction with cosmic
rays. We tested the sensitivity of Cold Electron Bolometers (CEBs) to ionizing radiation using gamma-rays from a radioactive source
and X-rays from a X-ray tube. We describe the test setup and the results. As expected, due to the effective thermal insulation of
the sensing element and its negligible volume, we find that CEBs are largely immune to this problem.


\end{abstract}

\section{Introduction}
The sensitivity of bolometers to cosmic rays is well known (see e.g. \cite{Caserta90}) and has been an important issue for several
space-based astronomy missions, including the recent Planck-HFI\cite{Planck13}. For future ultra-sensitive space-based surveys of
the sky in the mm/sub-mm range, like the proposed missions COrE\cite{Core11}, Millimetron, PRISM\cite{Prism13}, etc., which aim at
noise performance limited by the low photon background achievable in space, this will be the main factor limiting their ultimate
sensitivity (see e.g. \cite{Masi10}). In space one expects a mix of high energy  protons (with kinetic energy up to
1$\,$GeV\cite{Planck13}), neutrons and photons. For example, at balloon altitude, the typical fluxes are of the order of 1, 2, 30
m$^{-2}$s$^{-1}$ respectively. Cold Electron Bolometers (CEBs) represent a promising mm/sub-mm detection technology, in alternative
to the now common bolometers based on Transition Edge Sensors. In a CEB a nanoabsorber is coupled capacitively to the radiation
collecting antenna by means of SIN tunnel junctions. The same SIN junctions provide cooling of the nanoabsorber removing hot
electrons (see e.g. \cite{Kuzmin08a}). We have carried out a test campaign, irradiating CEBs built in Chalmers
\cite{Kuzmin08b,Perera12} using both gamma rays from radioactive sources and X-rays from an X-ray tube. We show below that the
results obtained in this way are also useful to estimate the effect of protons, taking into account the different spectra and
cross-sections of protons and photons. In the case of missions requiring large throughput detectors, like the SWIPE
instrument\cite{debe12} on the LSPE balloon\cite{LSPE12}, the effect of cosmic rays on standard bolometers can be very significant,
due to the large absorber area. Here we describe the experimental setup, the measurements and the results.

\section{Experimental setup}

\begin{figure} 
 \begin{minipage}[b]{5.5cm}
   \centering
   \includegraphics[width=0.95\linewidth,keepaspectratio]{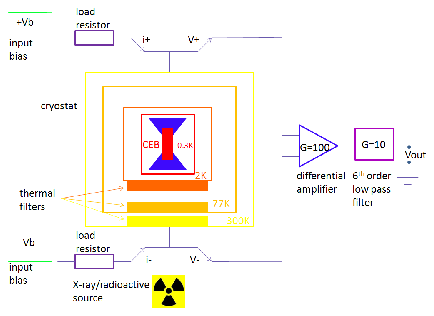}
   \caption{(Color online) Block diagram of the experimental setup for irradiation of
Cold Electron Bolometers with ionizing radiation.}
 \end{minipage}
 \ \hspace{0.1mm} \hspace{0.1mm} \
 \begin{minipage}[b]{5.5cm}
  \centering
   \includegraphics[width=0.80\linewidth,keepaspectratio]{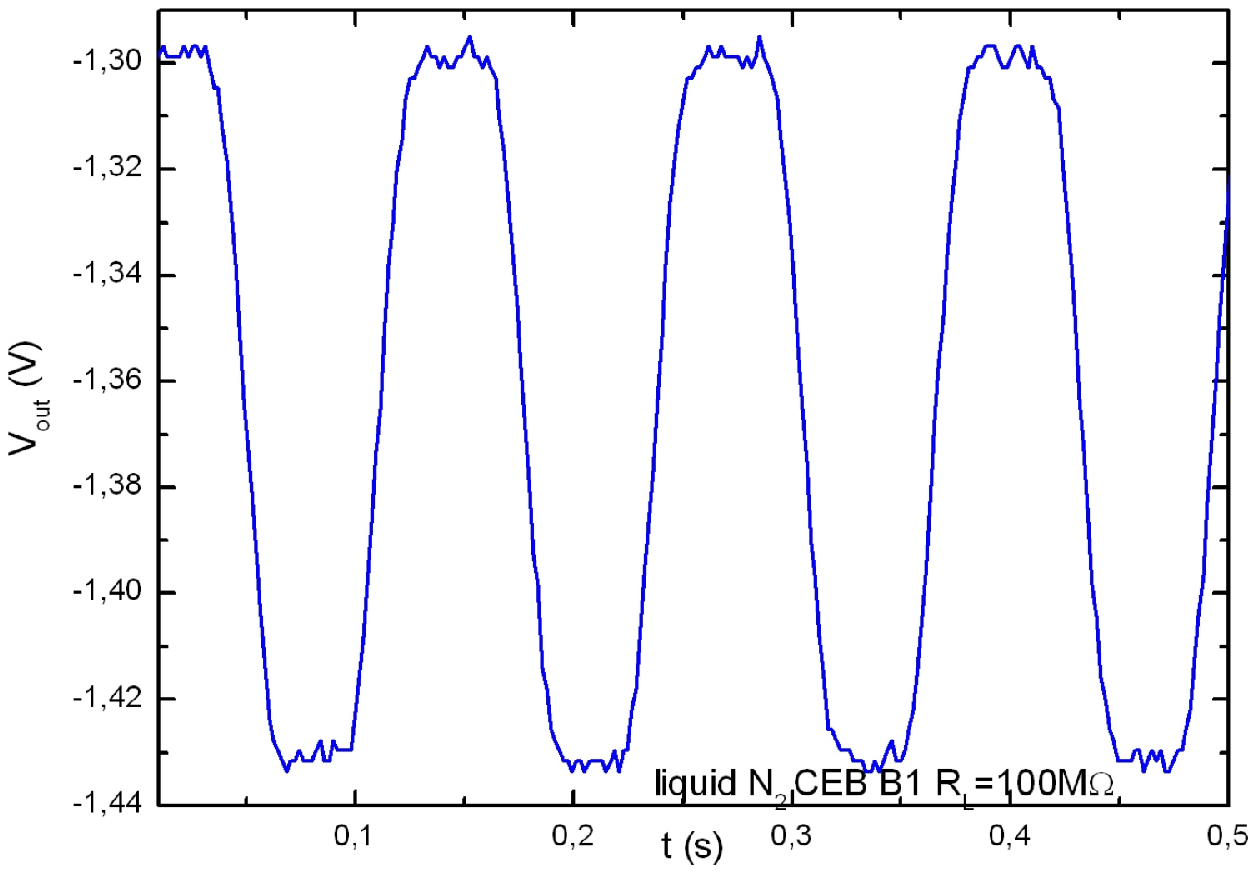}
   \caption{(Color online) Output voltage obtained chopping black-body radiation (300K$-$77K) in the 350$\,$GHz band of the detector.}
 \end{minipage}
\end{figure}

Due to the extremely small volume of the CEB absorber and to the relative decoupling of electron and phonon systems at low
temperatures, we expect that the CEB cross-section for ionizing particles is very small. We prepared our experimental setup to
check this hypothesis. The CEB is cooled down to about 304$\,$mK with a $^3$He fridge pre-cooled by a pulse tube refrigerator. A
window and a stack of filters defines the sensitive bandwidth of the detector (10$\%$ wide centered on 340$\,$GHz). The chip we
have tested couples to mm-wave photons through a small cross-slot antenna. The optical responsivity has been checked repeatedly
during the measurement campaign and found to be very stable. With optimal DC bias, the electrical responsivity is around
$2\times10^{7}\,$V/W. The detector signal is amplified by a factor 100 and filtered with a band pass filter (LF cut-off=0.1$\,$Hz,
HF cut-off=300$\,$Hz, gain=10, Sec.$\,$\ref{sec3}) or a 6th order low-pass filter (200$\,$Hz cut-off, gain=10, Sec.$\,$\ref{sec4}).
See Fig.$\,$1,2 for the setup and the response to mm waves. The rms fluctuation of the output signal in Sec.$\,$\ref{sec4} is of
the order of 3$\,$mV rms. This means that, at the detector, the noise level is 210 nV/$\sqrt{\textrm{Hz}}$. Using the responsivity
above, we find a NEP$\sim 2\times10^{-14}$W/$\sqrt{\textrm{Hz}}$. This NEP is significantly higher than the achievable NEP for this
kind of detectors. In our setup the dominant
\begin{figure}
 \begin{minipage}[b]{7cm}
   \centering
   \includegraphics[width=0.67\linewidth,keepaspectratio]{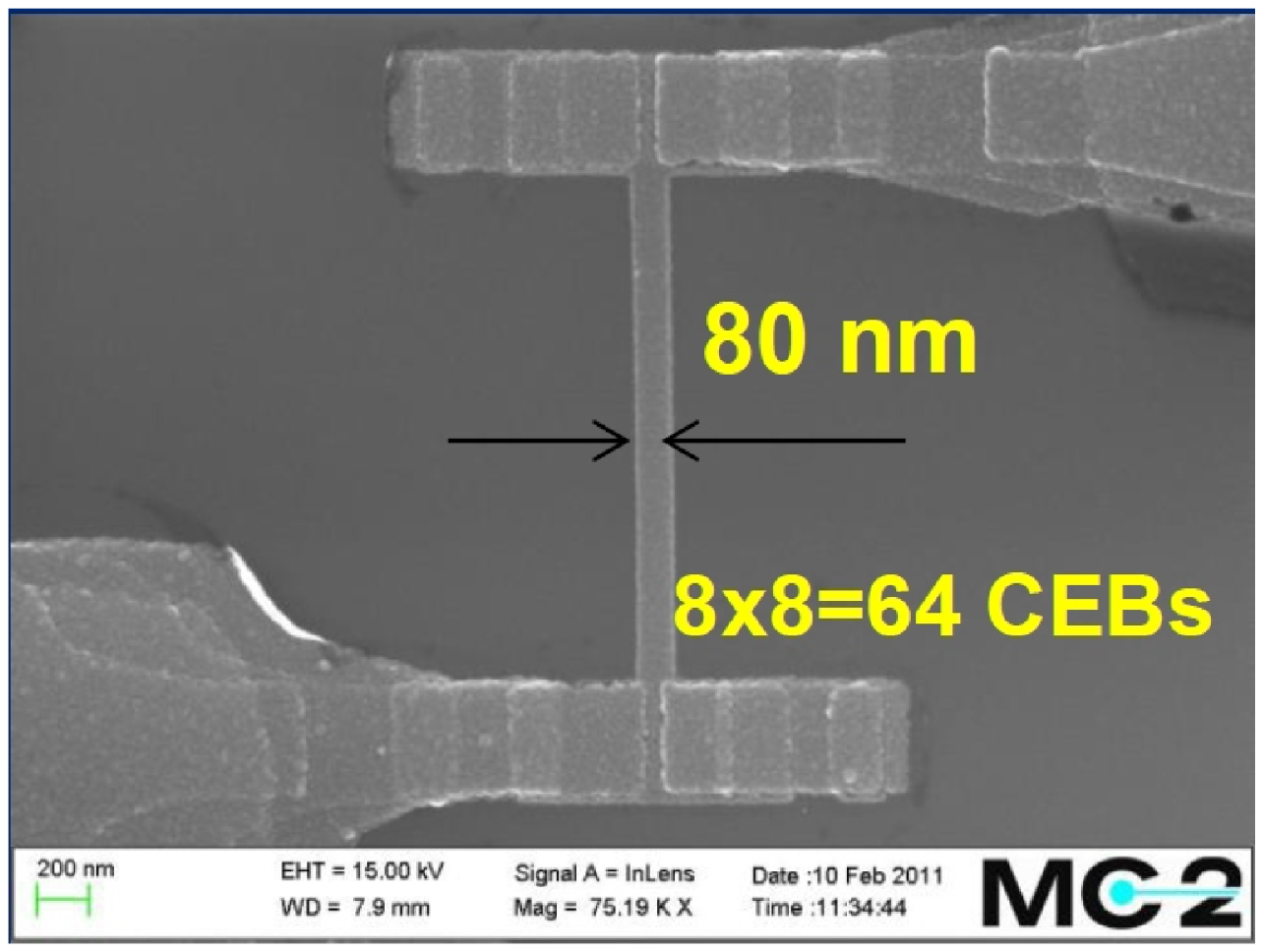}\label{fig1}
   \caption{(Color online) SEM picture of a typical CEB absorber.}
 \end{minipage}
 \ \hspace{0.1mm} \hspace{0.1mm} \
 \begin{minipage}[b]{4cm}
  \centering
   \includegraphics[width=0.58\linewidth,keepaspectratio]{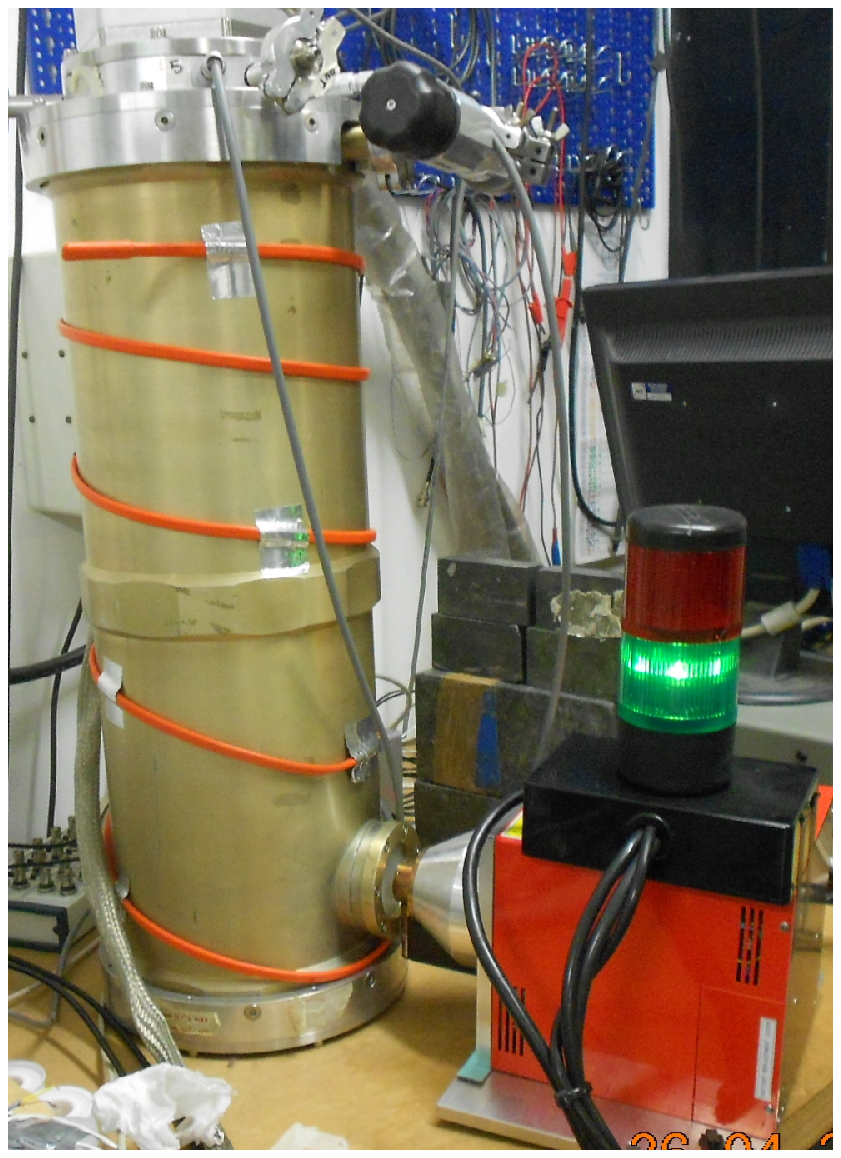}\label{fig2}
   \caption{(Color online) The microfocus X-ray source in front of the CEB cryostat.}
 \end{minipage}
\end{figure}
sources of noise are preamplifier noise (we did not have any cold JFET as an input stage) and the high photon background from the
300$\,$K laboratory. A source of ionizing photons is placed in front of the HDPE window of the cryostat. The (negligible)
absorption of ionizing photons by the window and the stack of filters is computed from literature data.

\section{Measurements with a radioactive source}\label{sec3}

We used a radioactive source made of the radionuclide $^{137}$Cs, with an activity 0.15$\,$MBq. (85.10$\pm$0.20)$\%$
of this emission consists of photons with an energy of (661.657$\pm\,\,$0.003)$\,$keV \cite{nndc}. From the geometry of our detector, the activity and distance of
the source (20.3$\,$cm), and the intervening absorption, we can compute the event of rates in two cases. If the entire CEB detector
area (4$\,$mm$^2$, silicon 280$\,\mu$m thick) is sensitive to ionizing particles we should observe one event every about 50$\,$s;
if only the Al absorbers (total area 5$\,\mu$m$^2$, thickness 10$\,$nm) are sensitive the events rate should be as low as about 1
event per month.

The noise power spectrum of V$_{out}$ does not change in presence of the radioactive source, nor its offset. For 662$\,$keV photons
the dominant interaction with the CEB is Compton scattering. Assuming that all the energy acquired by a target electron is
converted into a detectable signal, and taking into account the time response of our detection chain ($\sim0.4\,$ms), the signal
amplitude produced by each hit should be 1-4 mV at the detector; given the amplification of the readout electronics, it should be
easily detectable. We collected more than 16 hours of measurements finding none of such events. We conclude that either the only
part of the CEB chip sensitive to gamma-rays is the tiny CEB absorber (Fig.$\,$3), or the energy acquired by target electrons is
not converted into a detectable signal.

\section{Measurements with a X-ray source}\label{sec4}

\begin{figure}
\begin{center}
\includegraphics[%
  width=0.51\linewidth,
  keepaspectratio]{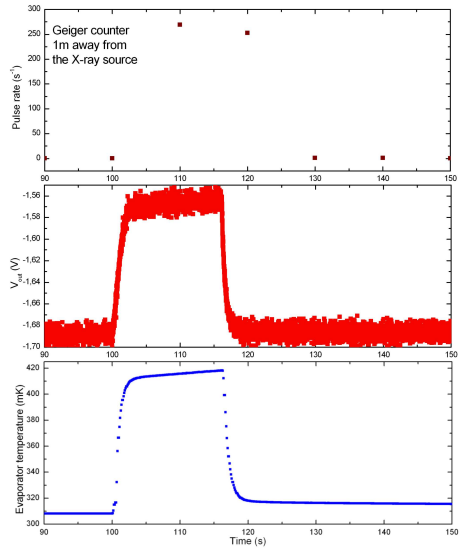}
\end{center}
\caption{(Color online) The effect of a large flux of X-ray
photons on a CEB. {\bf Top:} Record of a Geiger counter 1m away
from the X-ray source during the tests; the increase in the count
rate corresponds to source activity. {\bf Center:} Voltage at the
output of the CEB readout (V$_{out}$) in the same period, under
maximum source power (10$\,$W). {\bf Bottom:} Warm-up (!) of the
$^3$He evaporator in the same period. The recovery to the initial
temperature takes much longer than the recovery of the CEB
offset.} \label{g5}
\end{figure}

Having failed to detect ionizing particles with the radioactive source, we wanted to further check our hypothesis using a source of
ionizing particles producing a much higher flux, so that even if the sensitive volume is extremely small we should detect some
effect. We used a Microfocus X-ray source (Hamamatsu model L10101). We sent different fluxes of X-photons \cite{Klockenkämper97}
(Fig.$\,$4) in the energy range (10-100)$\,$keV. Spillover of X-rays was monitored by a Geiger counter 1m away from the X-ray
source (Fig.$\,$5, top). For large fluxes (high current in the source) and high energy (large accelerating voltage) (V $\times$ i
$>$ 2W) we observed a shift in the detector signal offset (Fig.$\,$5, center) and a heating of the $^3$He evaporator (Fig.$\,$5,
bottom). Both the heating of the evaporator and the offset shift are proportional to the integral of the Kramers' law over the
X-ray energies (Fig.$\,$6).

\begin{figure}
\begin{center}
\includegraphics[%
  width=0.9\linewidth,
  keepaspectratio]{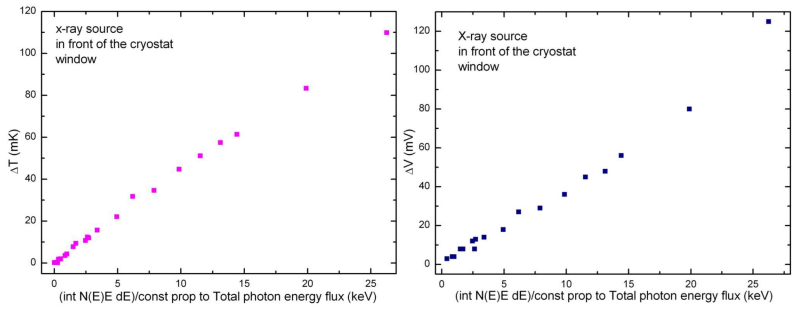}
\end{center}
\caption{(Color online) {\bf Left:} Cryostat evaporator temperature increase versus integrated continuum energy spectrum of X-rays
emitted by the X-ray source. {\bf Right:} V$_{out}$ offset shift versus the integrated continuum energy spectrum of X-rays emitted
by the X-ray source.} \label{g6}
\end{figure}

From the data of Fig.$\,$5 it is evident that the arrival of a
large number of X-ray photons per unit time results in a shift of
the detector signal offset, without any significant change of its
noise level. Either the temperature change of the evaporator
produces the change in the offset, or each single X-ray hit
produces a spike smaller than the instantaneous noise and the
offset change results as an integrated effect of many small
spikes. A combination of both effects is also possible. We note,
however, that the rms of the signal, both before and after
irradiation (detector and electronics noise only), and during the
irradiation (detector and electronic noise plus X-rays hits), is
very similar, with standard deviation around 3$\,$mV.

We can estimate the expected voltage signal produced by a X-ray photon hit on the CEB as follows. Assuming that all photons are
emitted at the wavelength of maximum luminosity, with an efficiency of 1$\%$\cite{Compton63}, we get a flux of about $10^{13}$
photons/s. Over a solid angle 42$^{\circ}$ FWHM wide and at the distance of our detector, we have $\dot{N}
=7\times10^{10}\textrm{s}^{-1}\textrm{cm}^{-2}$.

The expected event rate on the detector is given by $\dot{N}\,A\,P$, where $A$ is the area sensitive to energy deposition,
and $P$ is the interaction probability.
As before, we studied two cases: photons interactions with the CEBs absorber and with the CEBs area sensitive to the microwaves
and the corresponding substrate. In the
first case we should have one event every 12 minutes, while in the second case we should have many events for a single time
constant. The typical energy loss of a 50$\,$keV photon by Compton interaction is 1.5$\,$keV. So in the two cases we obtain an
expected power transferred to the detector of the order of 0.3$\,$pW for the absorber and 6$\,$nW for the detector area and
substrate. Given an electrical responsivity of 2$\times10^7\,$V/W, the expected signal, after amplification, should be 6$\,$mV for
the absorber and 130$\,$V for the whole detector area. The lack of saturation of our detector demonstrates that the area of the
detector chip is efficiently insulated from the absorber, so that the energy deposited elsewhere (e.g. in the substrate) is not
transferred to the absorber. The 6$\,$mV spikes due to energy lost directly in the absorber are not easy to separate from a 3$\,$mV
rms noise. Moreover, from the Kramers' law we expect that most of the spikes are of smaller amplitude. Also, this estimate assumes
that the energy lost by an X-ray photon is entirely converted into a useful signal in the absorber. We operated the X-ray tube for
a total time of about 14 minutes at different current and voltages, of which only 10$\,$s at the maximum current and voltage. So it
is not surprising that we were unable to detect any of these spikes. We can safely conclude that, at the NEP level we
operate, our detector is effectively immune to X and gamma rays.

Despite of the fact that these results have been obtained using X- and gamma- rays, we believe they are relevant for cosmic rays as
well. In fact the energy deposition of protons is maximum for 100$\,$keV protons, resulting in 1.2$\,$keV deposit (similar to the
energy deposition we have tested here with photons) and decreases for higher energy protons: a 1$\,$GeV proton
would deposit only 5$\,$eV as can be demonstrated using the proper protons stopping power data\cite{pstar}. So we can safely
conclude that the lack of detected events and the lack of noise increase using X-rays implies that the same will be true under
irradiation with cosmic rays in space.

This conclusion is corroborated by the fact that in space conditions\cite{Planck13}, the flux of ionizing particles will be many
orders of magnitude lower than in this experiment (about 5$\,\textrm{cm}^{-2}\textrm{s}^{-1}$). Having demonstrated that only the
tiny absorber area is sensitive to ionizing particles, this means that these detectors in space will be effectively immune from
cosmic rays hits.
\section{Conclusions}

We have tested the sensitivity of CEBs to ionizing radiation using photons from a radioactive source and an X-ray tube. We have
confirmed that the sensitive area is only the CEB absorber and not the entire detector area. We have also demonstrated that if
signal spikes are produced by X-rays, these are smaller than the rms noise of our detector, at a NEP level of $2\times
10^{-14}$W/$\sqrt{\textrm{Hz}}$. These experimental results confirm CEBs as very promising detectors to be used in future space
missions requiring ultra-sensitive mm to IR detectors.

\begin{acknowledgements}
The authors wish to thank Dr. D. Fargion  and Dr. I. Dafinei for
allowing us to use some of their instruments for our measurements.
This research has been funded in Italy by the Italian Space Agency
(grant I/022/11/0 LSPE).

\end{acknowledgements}


\end{document}